\documentclass[12pt,preprint]{aastex}

\newcommand{\x}{$\times$}

\slugcomment{Submitted to AJ}

\shorttitle{Cataclysmics in NGC 6397}
\shortauthors{Shara et al.}

\begin{document}

\title{Erupting Cataclysmic Variable Stars in the Nearest Globular
Cluster, NGC 6397: Intermediate Polars?\footnotemark[1]}

\footnotetext[1]{Based on observations with the NASA/ESA {\it Hubble Space
Telescope}, obtained from the Data Archive at the Space Telescope Science
Institute, which is operated by the Association of Universities for
Research in Astronomy, Inc., under NASA contract NAS5-26555. These
observations are associated with program \#8630.}

\author{Michael M. Shara}
\affil{Department of Astrophysics, American Museum of Natural History,
79th St. and Central Park West, New York, NY, 10024}
\email{mshara@amnh.org}

\author{Sasha Hinkley}
\affil{Department of Astrophysics, American Museum of Natural History,
79th St. and Central Park West, New York, NY, 10024}
\email{shinkley@amnh.org}

\author{David R. Zurek}
\affil{Department of Astrophysics, American Museum of Natural History,
79th St. and Central Park West, New York, NY, 10024}
\email{dzurek@amnh.org }

\author{Christian Knigge}
\affil{School of Physics and Astronomy, University of Southampton,
Southampton SO17 1BJ, United Kingdom}
\email{christian@astro.soton.ac.uk}

\author{Andrea Dieball}
\affil{School of Physics and Astronomy, University of Southampton,
Southampton SO17 1BJ, United Kingdom}
\email{andrea@astro.soton.ac.uk}


\begin{abstract}
NGC 6397 is the closest globular cluster, and hence the ideal place to
search for faint stellar populations such as cataclysmic
variables (CVs). HST and Chandra observers have identified nine certain
and likely CVs in this nearby cluster, including several magnetic CV
candidates. We have combined our recent UV imagery with archival HST
images of NGC 6397 to search for new CV candidates and especially to
look for dwarf nova-like eruptive events. We find remarkable and
somewhat unexpected dwarf nova-like eruptions of the two well-known
cataclysmic systems CV2 and CV3. These two objects have been claimed
to be {\it magnetic} CVs, as indicated by their helium emission-line
spectra.
Magnetic fields in CVs are usually expected to prevent the disk
instability that leads to dwarf nova eruptions. In fact, most field
magnetic CVs are observed to {\it not} undergo eruptions. Our
observations of the dwarf nova eruptions of CV2 and CV3 can be
reconciled with these objects' HeII emission lines if both objects are
infrequently-erupting intermediate polars, similar to EX Hya. If this is
the case for most globular cluster CVs then we can reconcile the many
X-ray and UV bright CV candidates seen by Chandra and HST with the very
small numbers of erupting dwarf novae observed in cluster cores.
\end{abstract}




\section{Introduction}

Galactic globular clusters are surprisingly rich in luminous X-ray
sources \citep{cla75,kat75}. Scenarios involving tidal capture
by two stars \citep{fab75} and/or exchange reactions involving three
stars \citep{hut83} are widely believed to be the sources of these strongly
interacting binaries in clusters. The remarkable correlation between
stellar encounter rate and number of X-ray sources in globular cluster
cores \citep{poo03} supports this viewpoint.

Accreting white dwarf-main sequence star binaries---the cataclysmic
variable (CV) stars---are almost certainly produced by the same
mechanisms in clusters, and should also be plentiful in globulars 
\citep{dis94}.
Perhaps the best cluster to search for this predicted population is NGC
6397. At a mere 3.3 kpc \citep{gra03} it is the closest globular cluster
to the Sun. Thus even CVs resembling the least luminous known such
objects (with $M \sim +11$) in this cluster should be detected with deep
HST imagery.

The initial discoveries of multiple X-ray sources \citep{coo93} and
H$\alpha$-bright stars \citep{coo95} in NGC 6397 were rapidly followed by
spectrographic identification of essentially certain CVs with hydrogen
and helium emission lines \citep{gri95,edm99}. Remarkably, all
four CVs in this cluster with HST spectra show significant, and in three
of four cases, prominent HeII 4686. This line is seen almost exclusively
in magnetic CVs and nova-like variables \citep{wil83, ech88},
prompting \citet{gri95} and \citet{gri99} to suggest that magnetic CVs
might well be the dominant CV population in globular clusters.

About half of all known field CVs \citep{dow01} are dwarf novae (DN).
Most field DN have been discovered because they undergo 2-5 magnitude
outbursts every few weeks to months \citep{war95}. In contrast,
strongly magnetic CVs are {\it not} observed to undergo DN outbursts.
This is because DN eruptions are likely caused by a disk instability
\citep{min83}, and the accretion disks of magnetic CVs have been truncated
or are absent.

Since large archival HST datasets of globular cluster images are
available, we have been systematically looking for erupting dwarf novae
in the cores of all such clusters \citep{sha96}.
Here we report the results of a search of NGC 6397 to determine
if any of the magnetic CV candidates undergo eruptions. Remarkably
we do find dwarf nova eruptions of two of the putative NGC 6397 magnetic
CVs.

\section{Observations}

The Hubble Space Telescope has imaged NGC 6397 during 5
separate epochs from 1996 to 2003. Our own 2003 datasets focused on STIS
UV imagery, heretofore unavailable and particularly useful in detecting
CVs. The dates of observation, PI and number of HST program, filters
used, number of frames and total exposure time in each filter are given
in the observing log which is Table 1.

\section{Multi-Wavelength Images and Light Curves of Two Dwarf Novae}

We have carried out visual inspection and aperture photometry of each of
the nine known CVs (referred to by \cite{gri01} as CV1 through CV9) in
every available archival HST image. While a modest level of variability
(up to a few tenths of a magnitude) is detectable in all nine CVs, and
two of the objects (CV1 and CV6) now have measured photometric periods
\citep{kal03}, none is yet reported to show a dwarf nova-like outburst.

The HST images of CV2 and CV3 are shown as photo-montages in Figures 1 and
2, respectively. The median brightness in every epoch, and in every
available filter is shown with mean filter wavelength running from blue
(left) to red (right). We also show the 5-epoch light-curves of these two
objects in Figure 3.

Each CV has been imaged in the WFPC2 F814W filter in epochs 1, 2 and 3,
and in the STIS CCDclear filter in epochs 4 and 5. We have normalized
epochs 1-3 with the first observation of each CV in epoch 1, and epochs
4-5 with the first observation in epoch 4.

The key result of this paper is that the putative magnetic cataclysmic
variables CV2 and CV3 are clearly detected as erupting dwarf novae.

CV2 is seen to brighten (in epoch 5) in the Space Telescope Imaging
Spectrograph (STIS) UV filter images. It is 2.7 magnitudes brighter than
it was in epoch 4.

CV3 is seen in eruption in the second epoch in visible and near infrared
pass-band images taken with the HST Wide Field and Planetary Camera
(WFPC2). It is 1.8 magnitudes brighter in eruption than in quiescence.
It is also seen in an intermediate brightness state (between
eruption and quiescence) in blue and near-UV WFPC2 images in epoch 3.
The rise time of CV3 in epoch 2 (about 2 days) is well in accord with
field dwarf nova rise times.

\section{Are the Erupting Systems Magnetic?}

The HST spectra of CV1, CV2 and CV3 are presented in \cite{gri95}; that
of CV4 is given in \cite{edm99}. HeII emission with equivalent widths in
the range 6-15 \AA \ appear in the spectra of all four objects, as
do strong Balmer lines. In particular, \cite{edm99} found for CV2 and CV3,
respectively, HeII 4686 equivalent widths of 11 and 15 \AA. The
H$\beta$ equivalent widths of these same two objects are 32 and 59 \AA,
respectively.

The disks in all four objects are faint ($M \sim 8-10$). \cite{edm99} has
carefully compared the disks and HeII line ratios of CVs 1-4 with those
of other cataclysmic systems. We refer the interested reader to that
careful and exhaustive analysis, simply quoting here Edmond et al.'s
conclusions regarding CVs 1-3:

``They do not appear to be recent novae or nova-likes because of their
faint disks (with extra evidence from their HeII 4686 \AA \ line ratios),
nor do they appear to be dwarf novae because they have moderately strong
HeII lines. The final option is magnetic systems. To conclude, CVs 1-3
do not appear to be dwarf novae, but they could be DQ Her-type
systems. A possible alternative to the DQ Her hypothesis is that some of
the NGC 6397 CVs are old novae (possibly in deep hibernation between
outbursts; see Shara et al. 1986).''

Figures 1 and 2 of this paper convincingly demonstrate that CV2 and CV3
are, in fact, erupting dwarf novae despite their prominent HeII emission.

Intensive monitoring with HST is essential to determine if all of the CV
candidates in NGC 6397---and other globular clusters---eventually erupt
as dwarf novae.

The hibernation scenario of CVs \citep{sha86} postulates that old novae
eventually display dwarf-nova like outbursts. One old classical nova in
the globular cluster M80 has been recovered \citep{sha95} at dwarf
nova-like luminosity, and one recent erupting classical nova in a
globular cluster (of the giant elliptical galaxy M87) has been reported
\citep{sha04}. CV2 and CV3 could be hibernating old novae, but we have,
at present, no observational proof that they once underwent nova
eruptions.

Edmonds' suggestion that the NGC 6397 CVs might be DQ Her-like magnetic
systems (also known as IPs or Intermediate Polars) might seem to be
disproved with the observed dwarf nova eruptions of CV2 and CV3.
However, there are a few well-documented cases of IPs that undergo dwarf
nova eruptions, including GK Per, EX Hya, XY Ari, DO Dra, TV Col, HT Cam
and V1223 Sgr.

Most dwarf novae show some HeII emission (see e.g. \cite{pat85}
for a compilation), so CV2 and CV3 are certainly not unique in this
respect. A comparison of the HeII and H$\beta$ line strengths for the CVs 
in NGC 6397
and for field IPs is presented by \cite{gri99}. Both the line strengths and
line ratios of HeII and H$\beta$ for CV2 and CV3 are comparable to
those of other low mass transfer rate IPs, with Ex Hya being a 
particularly good match.

\citet{ang89} have presented a model for dwarf nova-like outbursts in
magnetic CVs, which they apply to TV Col and GK Per. If the white dwarf
in a cataclysmic variable has a strong magnetic field, the inner region
of the accretion disk is disrupted. They found that disruption of the
inner disk region leads to shorter outbursts, and, for disk
instabilities starting near the inner disk edge, to much longer
intervals between outbursts.

The outbursts of most IPs tend to be rare, short and of only moderate
amplitude.
As an example, \cite{hel89} has noted that the EX Hya outbursts occur at
$\sim 2$ year intervals, last less than 4 days and are of only $\sim 3$ 
magnitude
amplitude. If the many globular cluster CV candidates identified by
Chandra \citep{ghe01} and HST \citep{kni02} are also intermediate polars 
then
the remarkably small numbers of erupting dwarf novae found by HST in
globular clusters \citep{sha96} can be explained. In particular, a
population of even dozens of IPs will only occasionally show a dwarf
nova near maximum light. Under this scenario we can predict that more of
these CVs will eventually be seen to erupt if individual clusters are
searched on dozens or hundreds of occasions.

\section{Conclusions}

We have observed NGC 6397 with HST during 5 separate epochs. Two of the
four spectrographically confirmed CVs are observed to undergo dwarf nova
eruptions with amplitudes of at least 1.8 and 2.7 magnitudes. These
eruptions were somewhat unexpected, as both objects display moderately
strong HeII 4686 emission lines, generally associated with non-erupting,
magnetic CVs. The dwarf nova-like eruptions reported in this paper can
occur in intermediate polars, suggesting that some of the NGC 6397 CVs
(and by implication some of the CVS in other globular clusters) are
indeed magnetic systems. The apparent rarity of erupting dwarf novae in
globular clusters with large CV candidate populations is reconciled if
those CVs are infrequently erupting intermediate polars.


\acknowledgments

Support for program \#8630 was provided by NASA through a grant from the
Space Telescope Science Institute, which is operated by the Association
of Universities for Research in Astronomy, Inc., under NASA contract NAS
5-26555."

\appendix


\clearpage


\begin{figure}
\figurenum{1}
\plotone{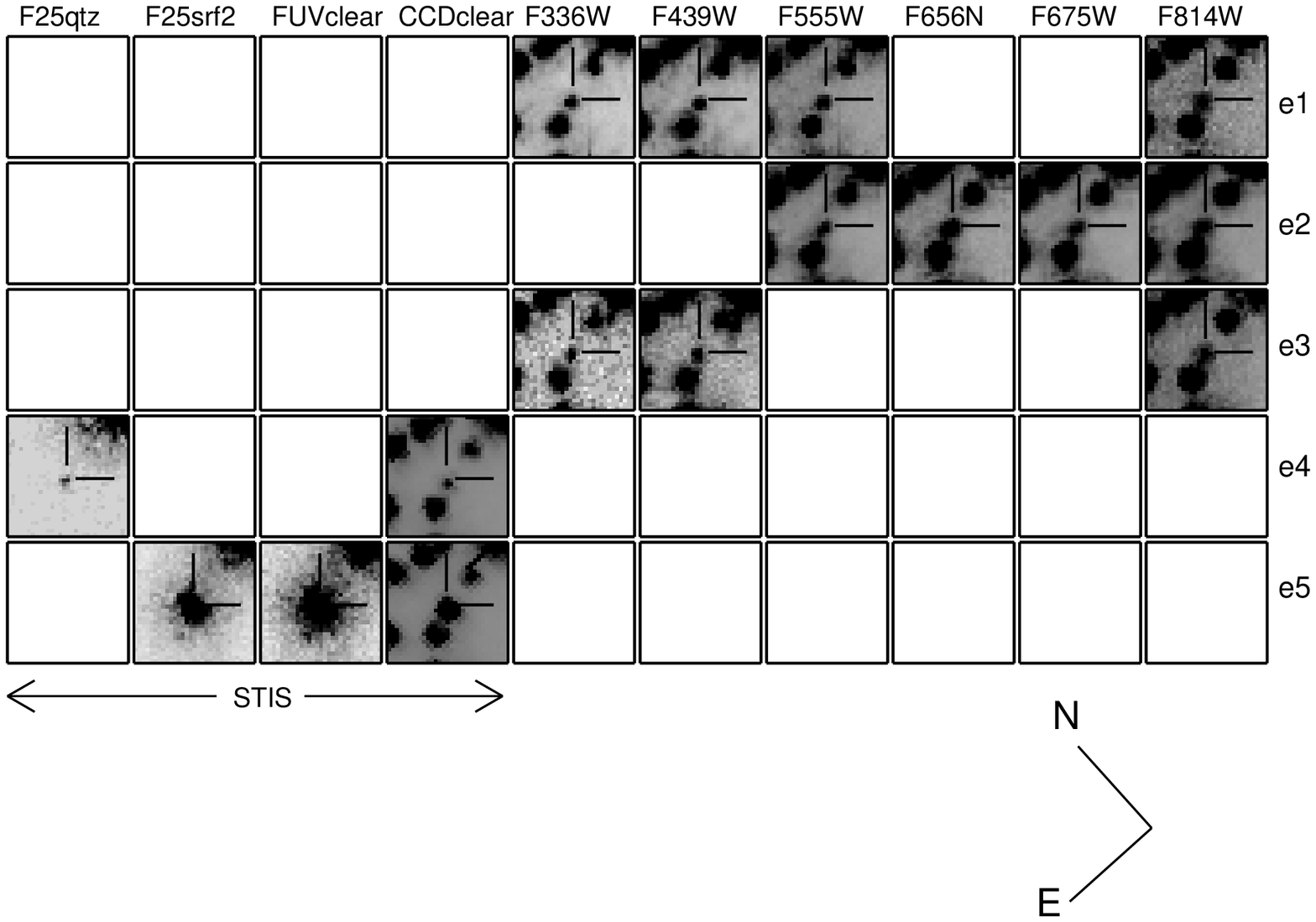}
\caption{Matrix of HST images of the field of CV2 in NGC6397. North and
East are as indicated, and each postage-stamp is 1.55'' on a side. The top
labels indicate the HST filter name, and are organized roughly in
increasing
wavelength from left (blue) to right (red). CV2 is indicated in each
image. }
\end{figure}

\begin{figure}
\figurenum{2}
\plotone{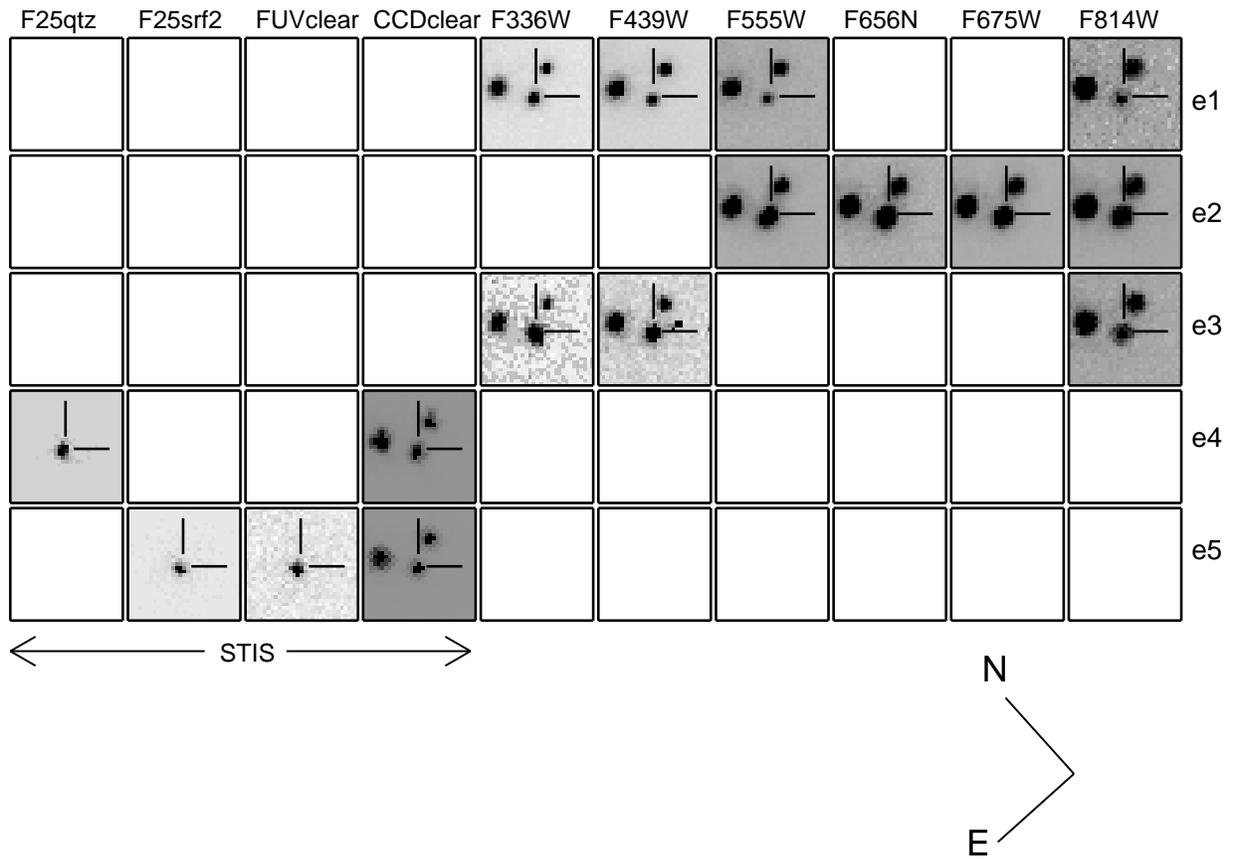}
\caption{Same as Fig. 1, but for CV3.}
\end{figure}

\begin{figure}
\figurenum{3}
\plotone{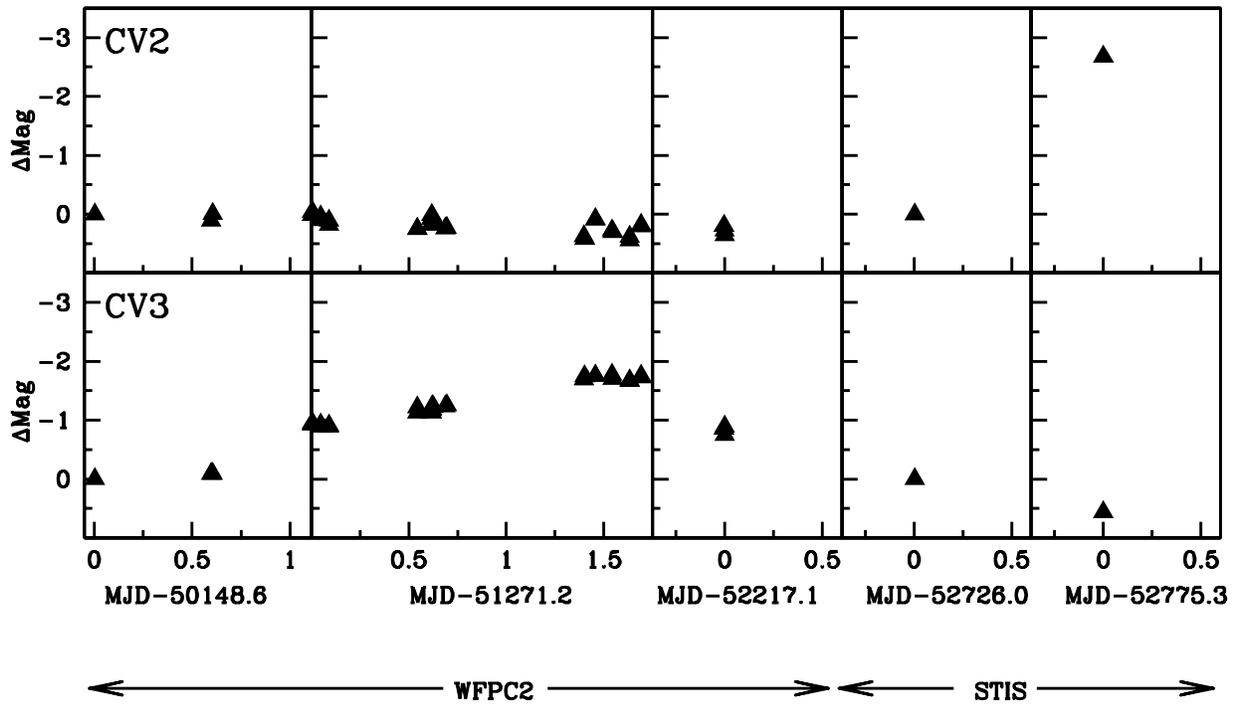}
\caption{A five-epoch lightcurve for CV2 (top) and CV3 (bottom). The first
three epochs show WFPC2 F814W photometry while the last two show the STIS
clear CCD data. The magnitudes are purely differential (see text), and the
MJDs are indicated for each epoch.}
\end{figure}

\begin{deluxetable}{l|l|c|c|c|c|c|c}
\tabletypesize{\scriptsize}
\tablecaption{Table of NGC6397 observations: Listed are the number of
useable observations at each date, as well as their exposure time.}
\tablewidth{0pt}
\tablehead{\colhead{WFPC2 Epochs} & \colhead{PI/Prog. ID} & \colhead{F336W} & \colhead{F439W} & \colhead{F555W} & \colhead{F656N} & \colhead{F675W} & \colhead{F814W}}
\startdata 
1:  3/6/96 & King        & 700s,6\x500s,   & 4\x500s,      & 6\x40s,8s,1s &                  &             &2\x40s,8s,1s \\   
           & \#5929      & 28\x400s,2\x80s,& 16\x400s,     &              &                  &             &             \\
           &             & 2\x10s          & 2\x80s,2\x10s &              &                  &             &             \\ \hline
2:  4/3/99 & Grindlay    &                 &               & 24\x40s,2\x8s& 1000s,4\x900s    & 38\x40s     &24\x40s      \\
           & \#7335      &                 &               & 2\x1s        & 18\x800s,7\x700s & 2\x8s,2\x1s &2\x8s,2\x1s  \\
           &             &                 &               &              & 2\x140s          &             &             \\ \hline
3: 11/4/01 & Noll        & 3\x160s         & 2\x160s       &              &                  &             & 3\x40s      \\
           & \#9313      &                 &               &              &                  &             &             \\\hline\hline
STIS Epochs& PI/Prog. ID &  Clear CCD      &  CCD LP       &  Clear MAMA  &  F25QTZ          &  F25SRF2    &             \\\hline\hline
4: 3/28/03 & Shara       & 3\x180s         & 3\x180s       &              & 533s,865s        &             &             \\
           & \#8630      &                 &               &              & 6\x600s          &             &             \\ \hline
5: 5/16/03 & ``''        & 3\x180s         & 3\x180s       & 3\x1000s     &                  & 2\x600s,883s&             \\
           &             &                 &               &              &                  & 2\x1021s    &             \\ \hline
\enddata
\end{deluxetable}

\end{document}